\documentclass[preprint,aps,showpacs]{revtex4-1}
\usepackage{geometry}                
\usepackage{amsmath,amssymb,amsfonts,dcolumn,color,graphicx,graphics,latexsym,placeins,epsfig,tikz}
\usetikzlibrary{arrows,shapes}
\usepackage{subfigure,rotating,bm,mathrsfs}

\newcommand{\be}{\begin{equation}}
\newcommand{\ee}{\end{equation}}
\newcommand{\ba}{\begin{eqnarray}}
\newcommand{\ea}{\end{eqnarray}}

\usepackage[pagebackref=false, colorlinks=true]{hyperref}
\definecolor{redish}{rgb}{0.7,0.2,0.0}  
\definecolor{bluish}{rgb}{0.2,0.5,0.8}

\hypersetup{linkcolor=redish,          
                  citecolor=blue,        
                  filecolor=magenta,      
                  urlcolor=blue}          

\begin{document}

\title{A novel gravitational lensing feature by wormholes}

\author{Rajibul Shaikh}
\email{rshaikh@iitk.ac.in}
\author{Pritam Banerjee}
\email{bpritam@iitk.ac.in}
\author{Suvankar Paul}
\email{svnkr@iitk.ac.in}
\author{Tapobrata Sarkar}
\email{tapo@iitk.ac.in}

\affiliation{{\it Department of Physics, Indian Institute of Technology, Kanpur 208016, India}}

\begin{abstract}
Horizonless compact objects with light rings (or photon spheres) are becoming increasingly popular in recent years for several reasons. In this paper, we show that a horizonless object such as a wormhole of Morris-Thorne type can have two photon spheres. In particular, we show that, in addition to the one present outside a wormhole throat, the throat can itself act as an effective photon sphere. Such wormholes exhibit two sets of relativistic Einstein ring systems formed due to strong gravitational lensing. We consider a previously obtained wormhole solution as a specific example. If such type of wormhole casts a shadow at all, then the inner set of the relativistic Einstein rings will form the outer bright edge of the shadow. Such a novel lensing feature might serve as a distinguishing feature between wormholes and other celestial objects as far as gravitational lensing is concerned. 

\end{abstract}


\maketitle

\section{Introduction}
Gravitational lensing is an established observational tool for probing gravitational fields around different compact objects. It is also used to test the viability of different alternative theories of gravity \citep{GTT}. After the theoretical prediction of light bending and its subsequent observational verification \citep{Dyson,Eddington}, interest in studying gravitational lensing has grown immensely in the last few decades. One of the fascinating features of gravitational lensing is that light can undergo an unboundedly large (i.e theoretically infinite) amount of bending in the presence of unstable light rings (or photon spheres in the spherically symmetric, static case) \citep{SL1,SL2,SL3,SL4}. As a result of such strong gravitational lensing, a large (theoretically infinite) number of relativistic images are formed.

In recent years, horizonless objects have attracted much attention for several reasons, one of them being that these objects with light rings can act as alternatives to black holes, i.e they can be ``black hole mimickers'' (see \citep{UCO1,UCO2,UCO3,UCO4,UCO5} and references therein). There has also been a lot of effort on how to distinguish such horizonless compact objects from black holes. For example, gravitational lensing and its various aspects, such as images, shadows etc by different horizonless objects such as wormholes \citep{WL1,WL2,WL3,WL4,WL5,WL6,WL7,WL8,WL9,WL10,WL11,WL12,WL13,WL14,WL15,WL16, WL17,WL18,WL19,WL20,WL21A,WL21}, naked singularities \citep{NL1,NL2,NL3,NL4,NL5,NL6,NL7,NL8}, Bosonic stars \citep{BSL1} etc. have been analyzed to see whether one can distinguish between these and a black hole. 
Such studies show that, in some cases, lensing by these objects are qualitatively different from that by a black hole. In this work, we study gravitational lensing by wormholes and point out a novel lensing feature which some wormholes can exhibit. In particular, we show that, in addition to a photon sphere present outside a wormhole throat, the throat can also act as an effective photon sphere. Such wormholes can exhibit interesting strong gravitational lensing features. In particular, they exhibit two sets of relativistic Einstein ring systems. To the best
of our knowledge, such a novel lensing feature has not been pointed out before in the literature. This is the main result of this paper that
we elaborate in sequel. 

The plan of the paper is as follows. In Sec. \ref{sec:NLF}, we work out the conditions under which the throat of a Morris-Thorne wormhole can act as an effective photon sphere, in addition to a photon sphere present outside the throat. We consider a specific example which satisfies all the energy conditions in a modified theory of gravity and study formation of relativistic images in Sec. \ref{sec:example}. We conclude Sec. \ref{sec:conclusion} with a summary and discussion of the key results.

\section{A novel gravitational lensing feature by wormholes}
\label{sec:NLF}
We consider a general spherically symmetric, static wormhole spacetime of the Morris-Thorne class. This spacetime is generically written as \citep{morris1}
\begin{equation}
ds^2=-e^{2\Phi(r)} dt^2+\frac{dr^2}{1-\frac{\mathcal{B}(r)}{r}}+r^2(d\theta ^2+\sin ^2\theta d\phi ^2),
\label{eq:general metric}
\end{equation}
where $\Phi(r)$ and $\mathcal{B}(r)$ are the redshift function and the wormhole shape function, respectively. The wormhole throat, where two different regions are connected, is given by $\left(1-\frac{\mathcal{B}(r)}{r}\right)\Big\vert_{r_0}=0$, i.e., by $\mathcal{B}(r_0)=r_0$, with $r_0$ being the radius of the throat. $\mathcal{B}(r)$ satisfies the flare-out condition $\mathcal{B}'(r_0)<1$ \citep{morris1}. The redshift function $\Phi(r)$ is finite everywhere (from the throat to spatial infinity).

For simplicity, we restrict ourselves to $\theta=\pi/2$. Because of the spherical 
symmetry, the same results can be applied to all $\theta$. Therefore, the 
Lagrangian describing the motion of a photon in the $\theta=\pi/2$ plane of 
the spacetime geometry \eqref{eq:general metric} is given by
\begin{equation}
2\mathcal{L}=-e^{2\Phi(r)}\dot{t}^2+\frac{\dot{r}^2}{1-\frac{\mathcal{B}(r)}{r}}+ r^2 \dot{\phi}^2,
\end{equation}
where a dot represents a derivative with respect to the affine
parameter. Since the Lagrangian is independent of $t$ and $\phi$, we
have two constants of motion given by 
\begin{equation}
p_t=\frac{\partial\mathcal{L}}{\partial\dot{t}}=-e^{2\Phi(r)}\dot{t}=-E~,~~
p_\phi=\frac{\partial\mathcal{L}}{\partial\dot{\phi}}=r^2\dot{\phi}=L~,
\end{equation}
where $E$ and $L$ are the energy and angular momentum
of the photon, respectively.  Using the null geodesics condition
$g_{\mu\nu}\dot{x}^\mu\dot{x}^\nu=0$, we obtain
\begin{equation}
\frac{e^{2\Phi(r)}}{1-\frac{\mathcal{B}(r)}{r}}\dot{r}^2+V_{eff}=E^2~, \hspace{0.3cm}
V_{eff}={L^2}\frac{e^{2\Phi(r)}}{r^2}~,
\label{eq:EP1}
\end{equation}
where $V_{eff}$ is the effective potential. From this, we can obtain the deflection angle given by \citep{SL3}
\begin{equation}
\alpha=2\int_{r_{tp}}^\infty\frac{e^{\Phi(r)}dr}{r^2\sqrt{1-\frac{\mathcal{B}(r)}{r}}\sqrt{\frac{1}{b^2}-\frac{e^{2\Phi(r)}}{r^2}}}-\pi~,
\end{equation}
where $b=L/E$ is the impact parameter, and $r_{tp}$ is the turning point given by $\frac{dr}{d\phi}=0$. $\frac{dr}{d\phi}=0$ gives the following 
relation between $b$ and $r_{tp}$:
\begin{equation}
b=r_{tp}e^{-\Phi(r_{tp})}.
\end{equation}
It is well-known that the deflection angle diverges logarithmically to infinity as the turning point approaches a photon sphere (which corresponds to the unstable photon orbits), and as a result, there are an infinite number of relativistic images formed due to lensing, just outside the photon sphere. Circular photon orbits satisfy $\dot{r}=0$ and $\ddot{r}=0$. Unstable photon orbits, which constitute the so-called photon sphere, additionally satisfy $\dddot{r}>0$. Generally, the conditions for the unstable photon orbit (photon sphere) are written in terms of the conditions on the effective potential. These are given by
\begin{equation}
V_{eff}(r_{ph})=E^2, \quad \frac{dV_{eff}}{dr}\Big\vert_{r=r_{ph}}=0, \quad \frac{d^2V_{eff}}{dr^2}\Big\vert_{r=r_{ph}}<0~,
\label{eq:PhSph_cond}
\end{equation}
where $r_{ph}$ is the radius of a photon sphere. Note that a photon sphere corresponds to the maximum of the potential. For the metric of 
Eq.(\ref{eq:general metric}), the above conditions become
\begin{equation}
b_{ph}=r_{ph}e^{-\Phi(r_{ph})}, \quad \Phi'(r_{ph})=\frac{1}{r_{ph}}, \quad \Phi''(r_{ph})< -\frac{1}{r_{ph}^2}~,
\label{eq:PhSph}
\end{equation}
where $b_{ph}$ is the critical impact parameter corresponding to the photon sphere. In addition to the photon sphere mentioned above, the effective potential can exhibit an extremum at the wormhole throat also, thereby implying circular photon orbits (either stable or unstable) at the throat. This can be seen by rewriting Eq. (\ref{eq:EP1}) as
\begin{equation}
\dot{r}^2+e^{-2\Phi(r)}\left(1-\frac{\mathcal{B}}{r}\right)\left(V_{eff}-E^2\right)=0~.
\end{equation}
Note that $\dot{r}=0$ is automatically satisfied at the wormhole throat $r=r_0$ since $\mathcal{B}(r_0)=r_0$ there. Therefore, the throat can act as an effective photon sphere when the other two conditions, namely $\ddot{r}=0$ and $\dddot{r}>0$, are satisfied. In terms of the effective potential, this becomes
\begin{equation}
V_{eff}(r_0)=E^2, \quad \frac{dV_{eff}}{dr}\Big\vert_{r=r_0}<0~,
\label{eq:EPhSph_cond}
\end{equation}
where we have used the flare-out condition $\mathcal{B}'(r_0)<1$. Note that the last set of conditions for the throat to be an effective photon sphere are different from those in (\ref{eq:PhSph_cond}) in the sense that the conditions in (\ref{eq:EPhSph_cond}) require one less derivative of the effective potential than those required for the photon sphere present outside the throat. This is due to the factor $\left(1-\frac{\mathcal{B}(r)}{r}\right)$ present in the radial geodesic equation. For the metric (\ref{eq:general metric}), Eq. (\ref{eq:EPhSph_cond}) becomes
\begin{equation}
b_{0}=r_0e^{-\Phi(r_0)}, \quad \Phi'(r_0)<\frac{1}{r_0},
\label{eq:EPhSph}
\end{equation}
where $b_0$ is the critical impact parameter corresponding to the effective photon sphere at the throat.

It is to be noted that, in the Schwarzschild gauge, the effective potential does not seem to have a maxima since $dV_{eff}/dr\neq 0$ at the throat, even though the throat acts as an effective photon sphere. This may seem confusing as traditionally the photon sphere is defined as the maximum of the effective potential. However, things become more clear when we switch over from the Schwarschild gauge to the proper radial coordinates defined by
\begin{equation}
l(r)=\pm\int_{r_0}^r\frac{dr}{\sqrt{1-\frac{\mathcal{B}}{r}}},
\end{equation}
where the throat is at $l(r_0)=0$, and the two signs correspond to the two different regions connected through the throat. In the proper radial coordinates, the radial geodesic equation becomes
\begin{equation}
e^{2\Phi(r(l))}\dot{l}^2+V_{eff}=E^2, \hspace{0.3cm}
V_{eff}={L^2}\frac{e^{2\Phi(r(l))}}{r^2(l)}.
\label{eq:EP2}
\end{equation}
Therefore, at the throat, we have
\begin{equation}
\frac{dV_{eff}}{dl}\Big\vert_{r_0}=\pm \sqrt{1-\frac{\mathcal{B}(r)}{r}}\frac{dV_{eff}}{dr}\Big\vert_{r_0}=0
\end{equation}
\begin{eqnarray}
\frac{d^2V_{eff}}{dl^2}\Big\vert_{r_0} = \left(1-\frac{\mathcal{B}(r)}{r}\right)\frac{d^2V_{eff}}{dr^2}\Big\vert_{r_0}+\frac{1}{2}\left(\frac{\mathcal{B}}{r^2}-\frac{\mathcal{B}'}{r}\right)\frac{dV_{eff}}{dr}\Big\vert_{r_0}
= \frac{(1-\mathcal{B}'(r_0))}{2r_0}\frac{dV_{eff}}{dr}\Big\vert_{r_0}
\end{eqnarray}
Now, using $\mathcal{B}'(r_0)<1$ and Eq. (\ref{eq:EPhSph_cond}), we obtain
\begin{equation}
V_{eff}(r_0)=E^2, \quad \frac{dV_{eff}}{dl}\Big\vert_{r_0}=0, \quad \frac{d^2V_{eff}}{dl^2}\Big\vert_{r_0}<0,
\end{equation}
which implies that, unlike in the Schwarzschild gauge (see Eq. (\ref{eq:EPhSph_cond})), in the proper radial coordinate, the effective potential has a maxima at the throat when the throat acts as an effective photon sphere.

Therefore, in addition to the outer photon sphere, the wormhole throat can also act as an effective photon sphere. As a result, two sets of infinite number of relativistic images may be formed due to strong gravitational lensing of light coming from a distant light source. However, for the effective photon sphere at the throat to take part in the formation of relativistic images, the potential must have higher height at the throat than at the photon sphere present outside the throat. Therefore, the necessary and sufficient conditions that the throat also takes part in the formation of relativistic images in addition to those formed due to the outer photon sphere are given by
\begin{equation}
\Phi'(r_0)<\frac{1}{r_0},  \quad \frac{e^{2\Phi(r_0)}}{r_0^2}>\frac{e^{2\Phi(r_{ph})}}{r_{ph}^2},
\label{eq:sufficient_cond}
\end{equation}
where $r_0$ and $r_{ph}$ are given by
\begin{equation}
\mathcal{B}(r_0)=r_0, \quad \Phi'(r_{ph})=\frac{1}{r_{ph}}.
\end{equation}
Note that there should be a minima between the throat and the outer photon sphere. Therefore, when the throat acts as an effective photon sphere, maxima and minima outside the throat must come in pairs, and the condition $\Phi''(r_{ph})<-\frac{1}{r_{ph}^2}$ will automatically be satisfied at the maxima.

Let us now show how we can construct a wormhole metric which possesses an effective photon sphere at its throat, as well as a photon sphere outside its throat. This demands that we must have $ V'_{eff} = 0 $ at two different radii outside the throat, with one of them representing the maxima at the outer photon sphere and the remaining one representing the minima of $ V_{eff} $ present in between the throat and the outer photon sphere. Let us consider the following ansatz for $ V'_{eff}(r) $ :
\begin{equation}
V'_{eff}(r) = -K \frac{(r - \beta)(r - \gamma)}{r^n} \label{4}
\end{equation}
where $ K $ is a positive constant, $ n $ is a positive integer and $ r_0 \le \beta < \gamma $. $ V'_{eff}(r) $ is zero at two points ($ \beta $, $ \gamma $) with $ \gamma $ representing maxima, i.e., $ V''_{eff}(r) \big| _{\gamma} < 0 $, and $ \beta $ representing minima of $ V_{eff}(r) $, i.e. $ V''_{eff}(r) \big|_\beta > 0 $. Integrating the above form of $ V'_{eff}(r) $, we obtain
\begin{eqnarray}
V_{eff}(r) = - K \int \frac{(r - \beta)(r - \gamma)}{r^n} ~ dr 
= K r^{-n} \left(\frac{r^3}{n-3}-\frac{r^2 (\beta +\gamma )}{n-2}+\frac{r (\beta  \gamma )}{n-1} \right) \label{5}
\end{eqnarray}
From Eqs. (\ref{eq:EP1}) and (\ref{5}), we find
\begin{equation}
e^{2 \phi(r)} = K r^{2-n} \left(\frac{r^3}{n-3}-\frac{r^2 (\beta +\gamma )}{n-2}+\frac{r (\beta  \gamma)}{n-1} \right) \label{6}
\end{equation}
where the constant of motion $ L $ is absorbed in $ K $.

Equation (\ref{6}) shows that $ e^{2 \phi(r)} $ remains finite in the range $ r \in \left[ r_0,\infty \right) $ for $ n \ge 5 $. Again, the asymptotically flat condition, i.e., $ e^{2 \phi(r)} \to 1 $ in the limit $ r \to \infty $, is satisfied only for $ n = 5 $. Putting this value of $ n $ in Eq. (\ref{6}) and taking the $ r \to \infty $ limit, we get
$$ \lim_{r\to\infty}e^{2 \phi(r)} = 1 \implies \frac{K}{2} = 1 \implies K=2 $$
Therefore, the final form of $ e^{2 \phi(r)} $ becomes
\begin{equation}
e^{2 \phi(r)} =  2 \left(\frac{1}{2}-\frac{(\beta +\gamma )}{3 r}+\frac{(\beta  \gamma)}{4 r^2}\right) \label{7}
\end{equation}
Now we should choose $\beta$ and $\gamma$ in such a way that the conditions in (\ref{eq:sufficient_cond}) (with $r_{ph}=\gamma$) are satisfied. Let us illustrate by taking a specific choice $\beta=\frac{3}{2}r_0$ and $\gamma=\frac{5}{2}r_0$. Also, we choose $\mathcal{B}(r)=\frac{8r_0}{3}-\frac{5r_0^2}{3 r}$ so that $\mathcal{B}(r_0)=r_0$, and $M=\frac{4}{3}r_0$ is the Arnowitt-Deser-Misner 
mass of the wormhole. Therefore, we have
\begin{equation}
e^{2 \phi(r)} =1-\frac{8r_0}{3r}+\frac{15r_0^2}{8 r^2}, \quad 1-\frac{\mathcal{B}(r)}{r}=1-\frac{8r_0}{3r}+\frac{5r_0^2}{3 r^2}.
\label{7}
\end{equation}
The above wormhole metric was constructed by using the fact that it must have an effective photon sphere at its throat as well as a photon sphere outside its throat. In the next section, however, we consider a wormhole example which possesses the above feature and arises as an exact solution in a modified theory of gravity.

\section{A specific example in modified gravity}
\label{sec:example}

We consider a wormhole solution obtained in \cite{rajibul_2018}. The spacetime geometry is given by
\begin{equation}
ds^2=-\psi^2(r)f(r)dt^2+\frac{dr^2}{f(r)}+r^2\left(d\theta^2+\sin^2\theta d\phi^2\right),
\end{equation}
\begin{equation}
\psi(r)=\frac{1}{\sqrt{1+\kappa \rho}}
\end{equation}
\begin{eqnarray}
f(r) &=& \frac{1+\kappa\rho}{1-\kappa p_{\theta}}\left[1-\sqrt{\frac{1-\frac{\kappa C_0}{2r^4}}{1+\frac{\kappa C_0}{2r^4}}}\left(\frac{2M}{r}-\frac{C_0}{r^2}{\;}_2F_1\left[\frac{1}{2},\frac{1}{8},\frac{9}{8};\frac{\kappa^2C_0^2}{4r^8} \right]\right. \right. \nonumber\\
& & \left. \left. +\frac{\kappa C_0^2}{30r^6}{\;}_2F_1\left[\frac{1}{2},\frac{5}{8},\frac{13}{8};\frac{\kappa^2 C_0^2}{4r^8} \right]\right) -\frac{\kappa C_0^2}{6r^6\left(1-\frac{\kappa C_0}{2r^4}\right)}\right],
\label{eq:fr}
\end{eqnarray}
where $\rho$ is the energy density given by
\begin{equation}
\rho=\frac{C_0}{r^4}\frac{1}{1-\frac{\kappa C_0}{2r^4}},
\nonumber
\end{equation}
where $C_0$ is an integration constant. It is to be noted that the factor $\left(1-\frac{\kappa C_0}{2r^4}\right)$ present in the denominator of the last term in Eq. (\ref{eq:fr}) is absent in \cite{rajibul_2018}. This was a typographic error in that paper. For $\kappa<0$, the solution represents a wormhole, and the wormhole throat radius is given by $(1+\kappa\rho)\vert_{r_0}=0$, which gives $r_0=\left(\frac{|\kappa| C_0}{2}\right)^{1/4}$ \cite{rajibul_2018}. Additionally, we must have $x=\frac{r_0^2}{|\kappa|}<1$ for the wormhole. For $x>1$, the throat is covered by an event horizon, thereby giving a black hole solution. The matter supporting the above wormhole satisfies all the energy conditions. However, to make the wormhole traversable, the parameter $\kappa$, the mass $M$ and the throat radius $r_0$ of the wormhole have to satisfy the following \citep{rajibul_2018}:
\begin{equation}
M=\frac{r_0^3}{|\kappa|} {\;}_2F_1\left[\frac{1}{2},\frac{1}{8},\frac{9}{8};1 \right]+\frac{r_0^3}{15|\kappa|} {\;}_2F_1\left[\frac{1}{2},\frac{5}{8},\frac{13}{8};1 \right].
\end{equation}

\begin{figure}[ht]
\centering
\subfigure[$x=0.6$]{\includegraphics[scale=0.55]{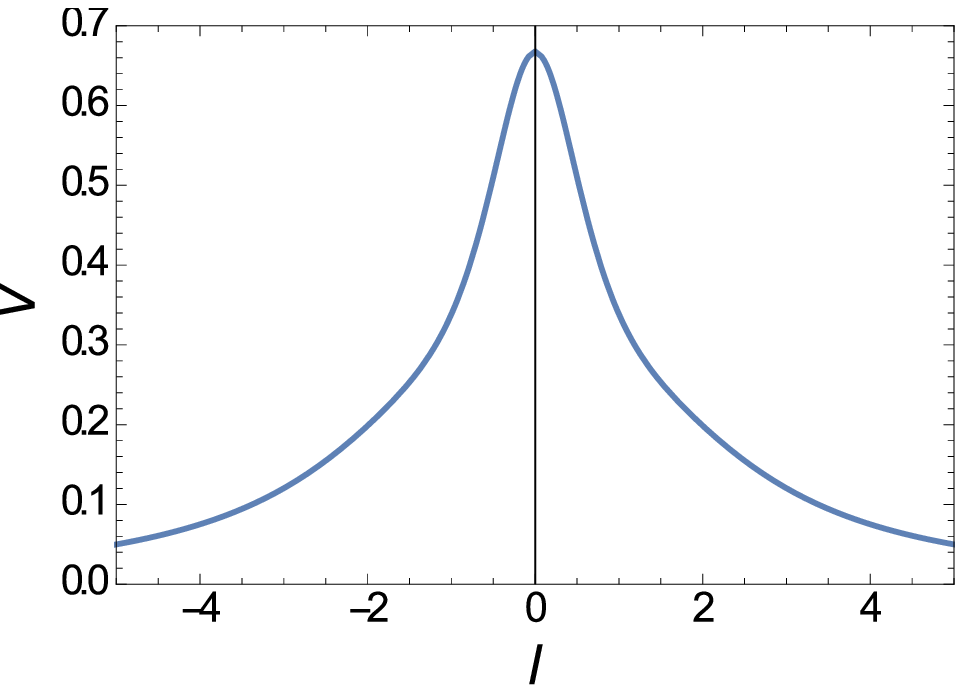}\label{fig:pot1}}
\subfigure[$x=0.85$]{\includegraphics[scale=0.55]{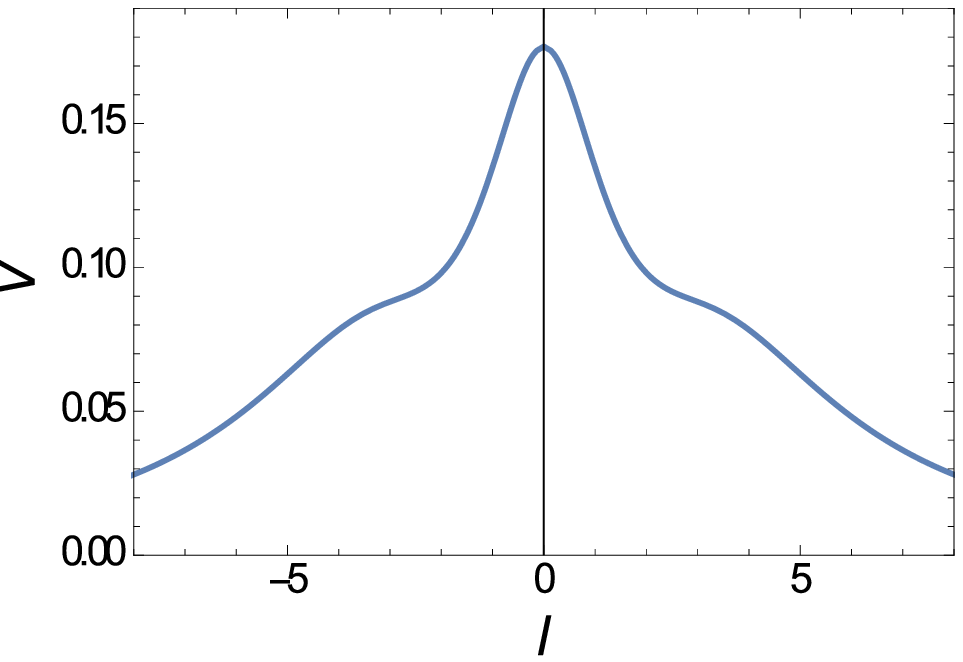}\label{fig:pot3}}
\subfigure[$x=0.9$]{\includegraphics[scale=0.55]{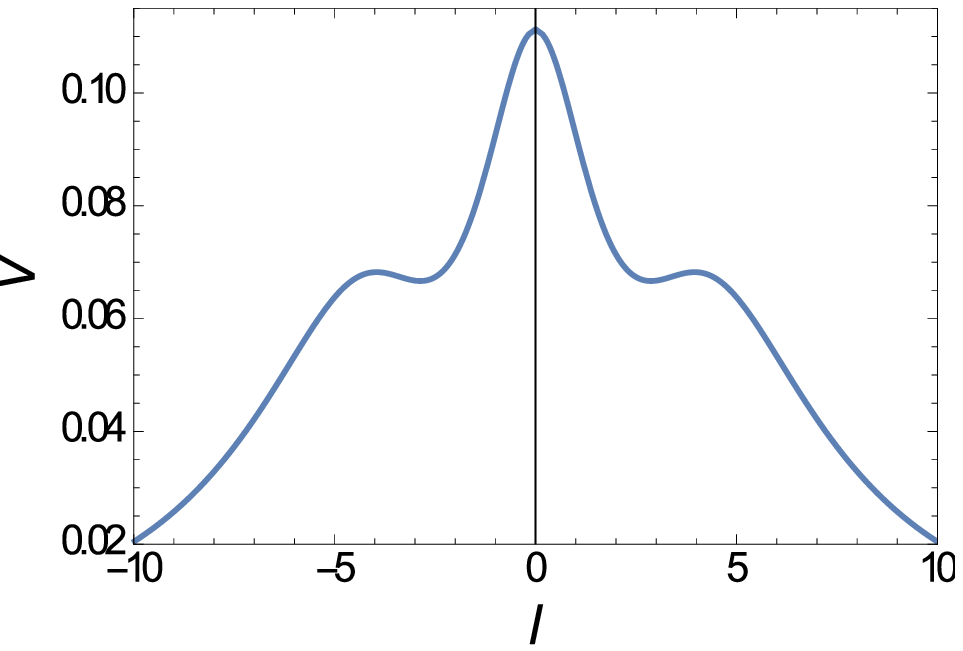}\label{fig:pot4}}
\subfigure[$x=0.93$]{\includegraphics[scale=0.55]{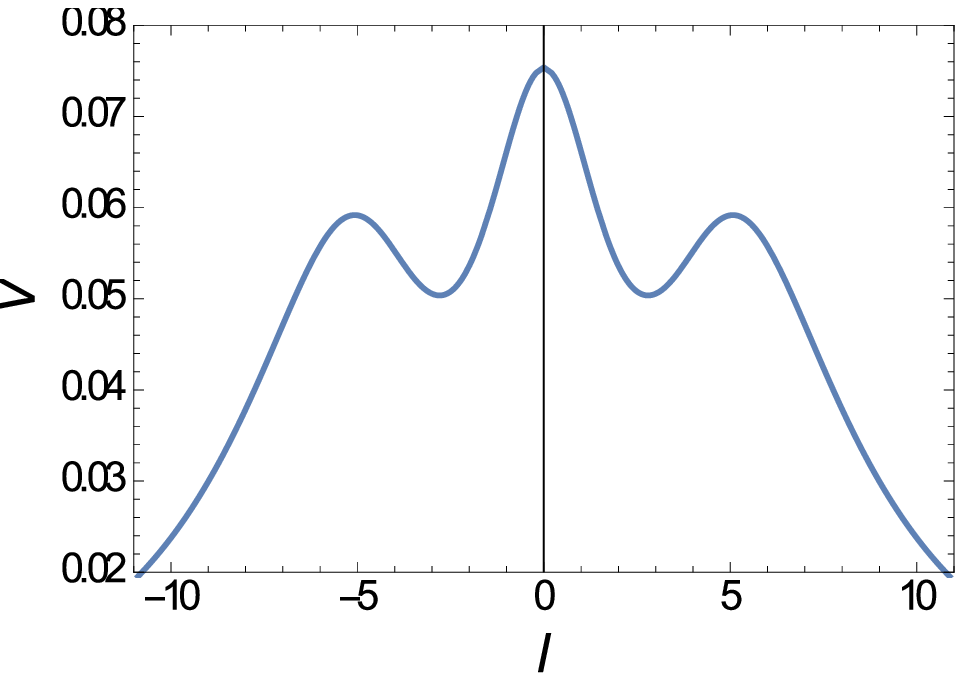}\label{fig:pot5}}
\subfigure[$x=0.96$]{\includegraphics[scale=0.55]{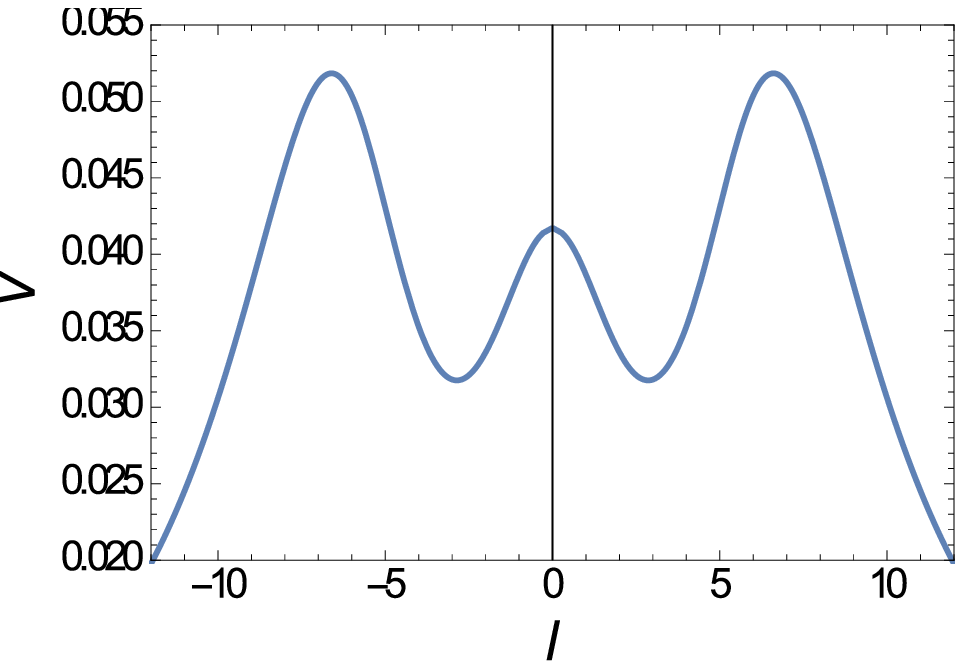}\label{fig:pot6}}
\subfigure[$x=1.1$]{\includegraphics[scale=0.55]{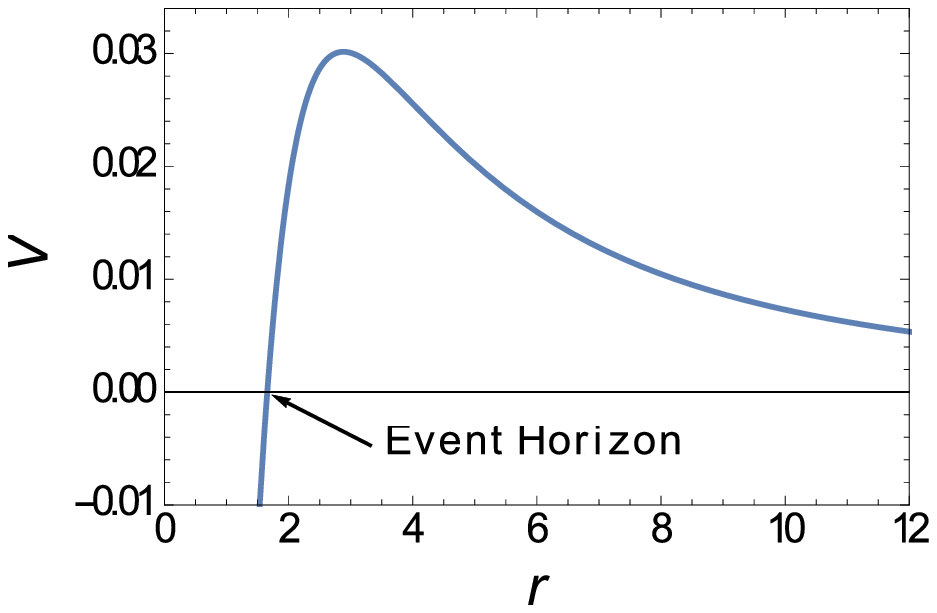}\label{fig:pot7}}
\caption{The effective potential $V$ ($=\frac{V_{eff}}{L^2}$) plotted for different $x$ ($=\frac{r_0^2}{|\kappa|}$) and $\kappa=-1$.}
\label{fig:pot}
\end{figure}

Figure \ref{fig:pot} shows the effective potential for different value of $x$ (both wormhole and black hole). In Figs. \ref{fig:pot1} and \ref{fig:pot3}, only the throat acts as an effective photon sphere. In Figs. \ref{fig:pot4}-\ref{fig:pot6}, there is an outer photon sphere in addition to the effective photon sphere at the throat. However, unlike the effective photon sphere at the throat in Figs. \ref{fig:pot4} and \ref{fig:pot5}, that in Fig. \ref{fig:pot6} will be hidden within the outer photon sphere and hence will not take part in the formations of relativistic images in gravitational lensing since the effective potential at the throat has lesser height than that at the outer photon sphere.

\begin{figure}[ht]
\centering
\subfigure[$x=0.6$]{\includegraphics[scale=0.55]{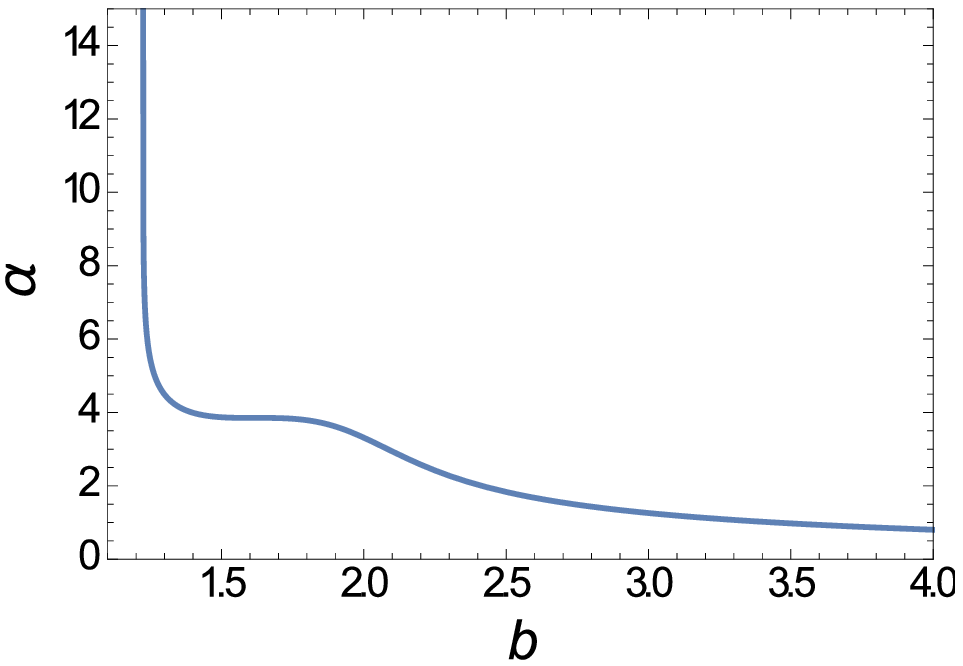}\label{fig:def1}}
\subfigure[$x=0.85$]{\includegraphics[scale=0.55]{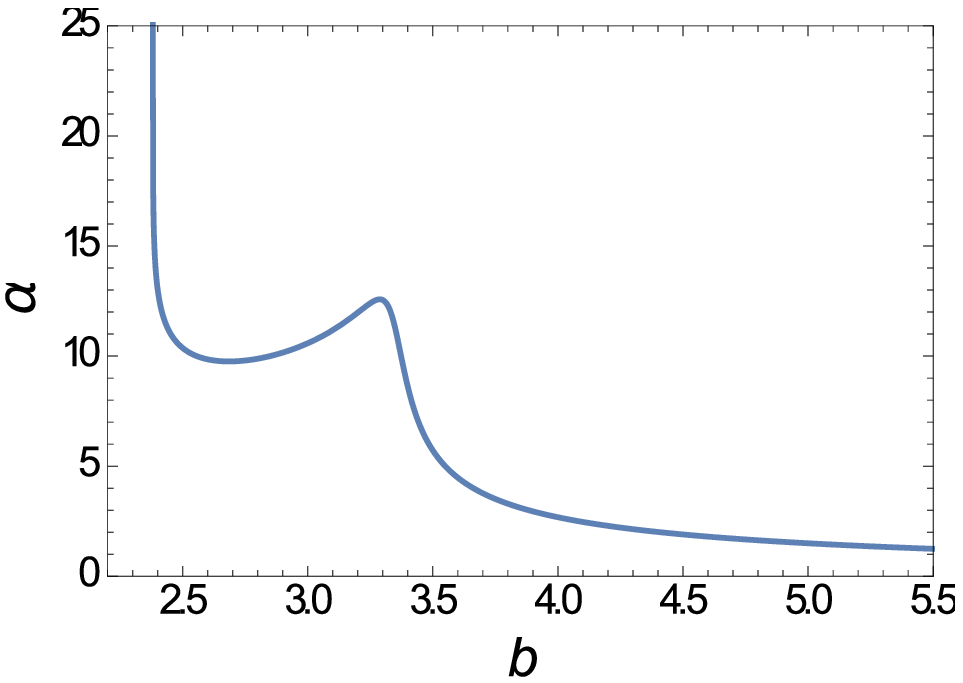}\label{fig:def3}}
\subfigure[$x=0.9$]{\includegraphics[scale=0.55]{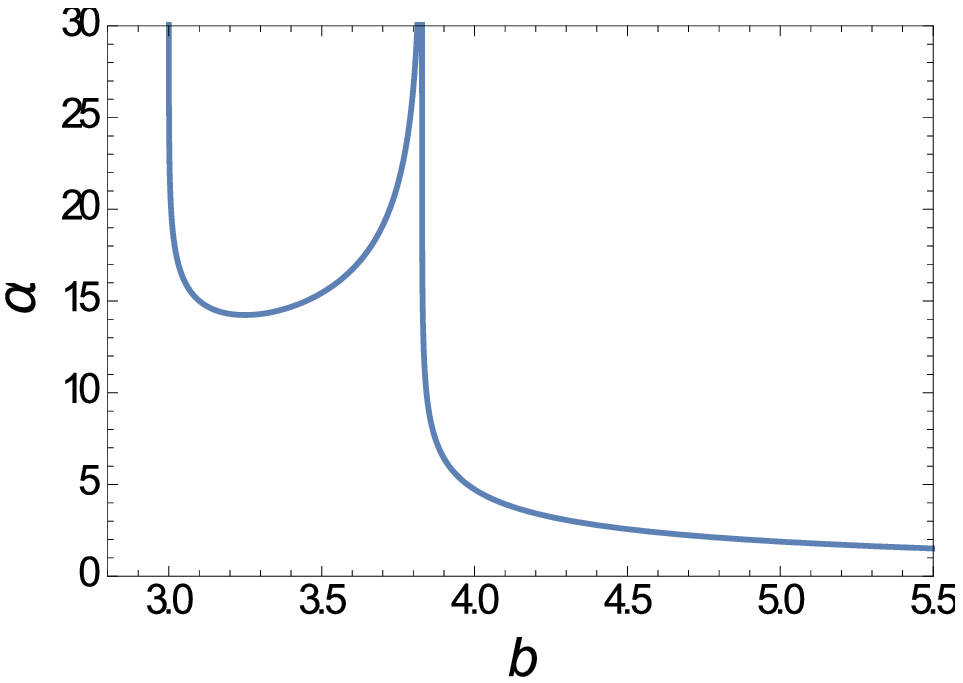}\label{fig:def4}}
\subfigure[$x=0.93$]{\includegraphics[scale=0.55]{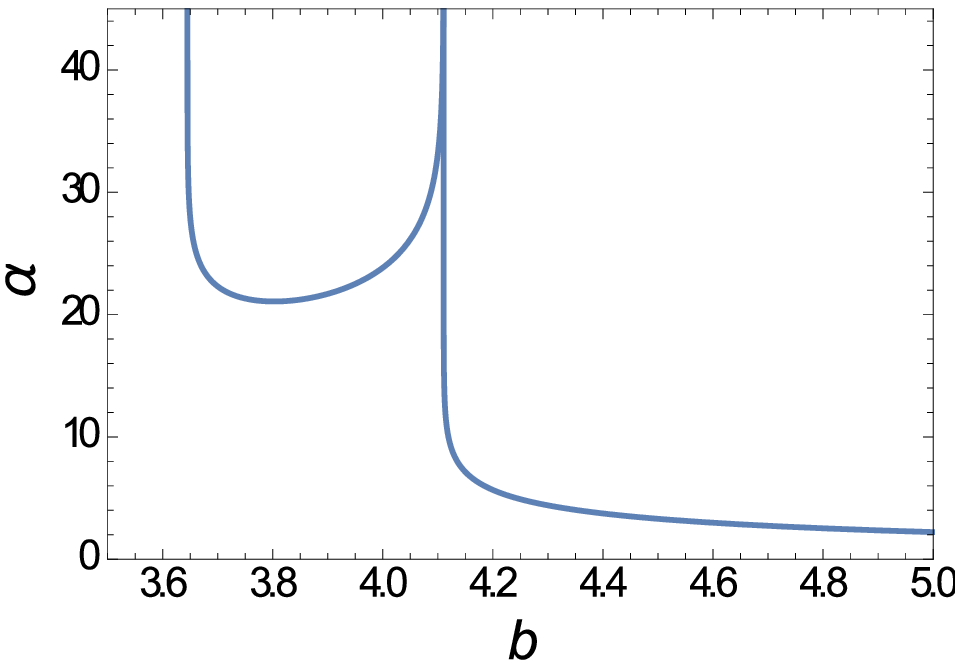}\label{fig:def5}}
\subfigure[$x=0.96$]{\includegraphics[scale=0.55]{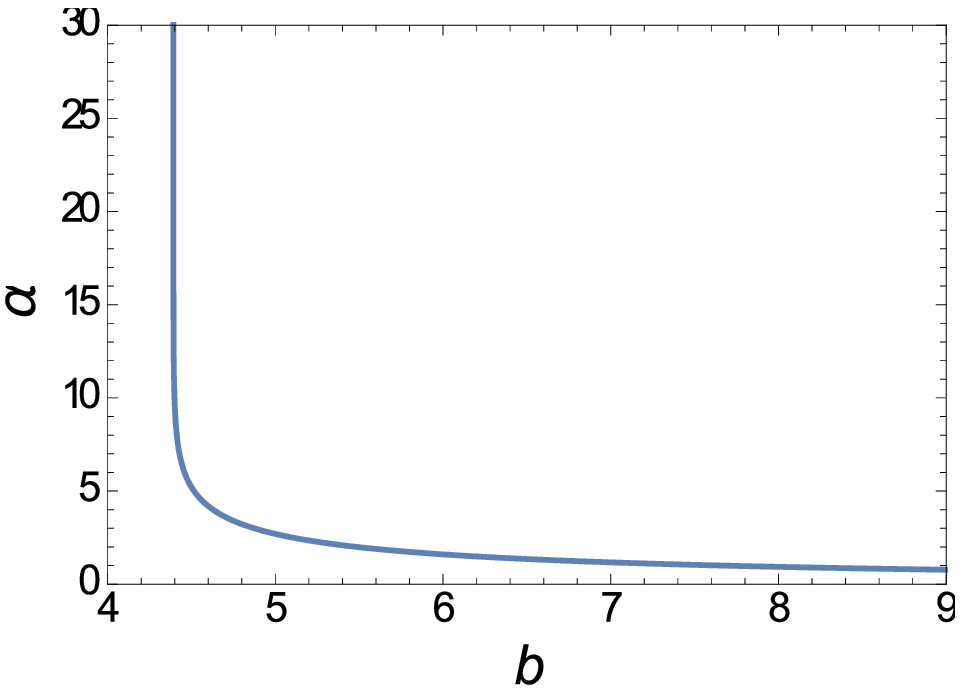}\label{fig:def6}}
\subfigure[$x=1.1$]{\includegraphics[scale=0.55]{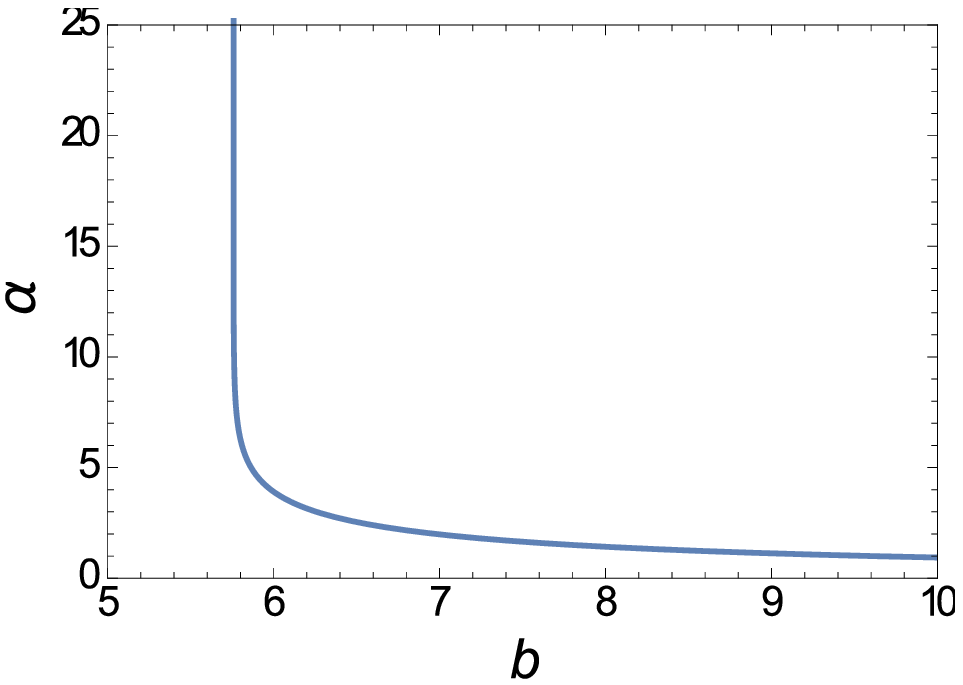}\label{fig:def7}}
\caption{The deflection angle plotted as a function of the impact parameter for different $x$ ($=\frac{r_0^2}{|\kappa|}$) and $\kappa=-1$.}
\label{fig:def}
\end{figure}

Figure \ref{fig:def} shows the numerically integrated deflection angle as a function of the impact parameter for the metric parameter values same as those in Fig. \ref{fig:pot} (both wormhole and black hole). Note that, as expected, the deflection angle diverges as the impact parameter approaches the values corresponding to the effective photon sphere at the throat or the outer photon sphere. In the case where there is both the effective photon sphere at the throat as well as the outer photon sphere, the deflection angle diverges at the locations of both the throat and the outer photon sphere. In this case, as we decrease the impact parameter, the deflection angle diverges to infinity as $b\to b_{ph}$. As we decrease $b$ further below $b_{ph}$, the deflection angle start decreasing, but is again divergent as $b$ approaches the critical value $b_0$ corresponding to that of the effective photon sphere at the throat.

\begin{figure}[ht]
\centering
\subfigure[$x=0.6$]{\includegraphics[scale=0.55]{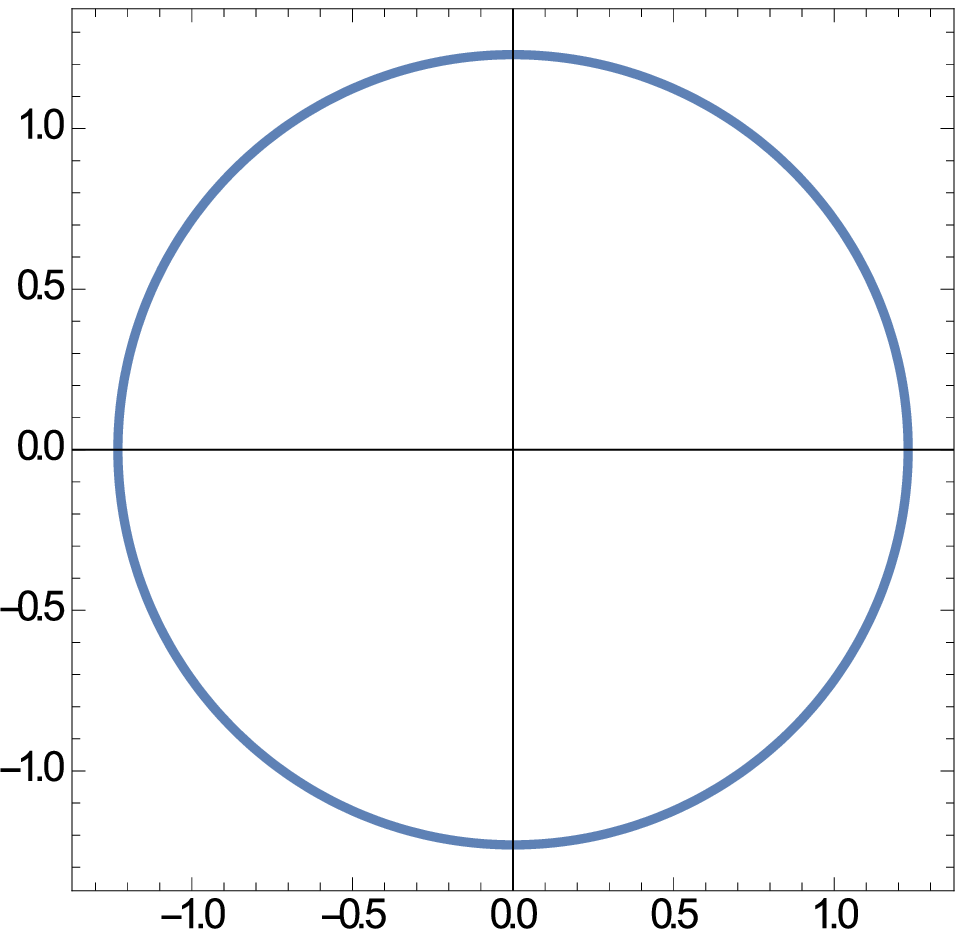}\label{fig:ring1}}
\subfigure[$x=0.85$]{\includegraphics[scale=0.55]{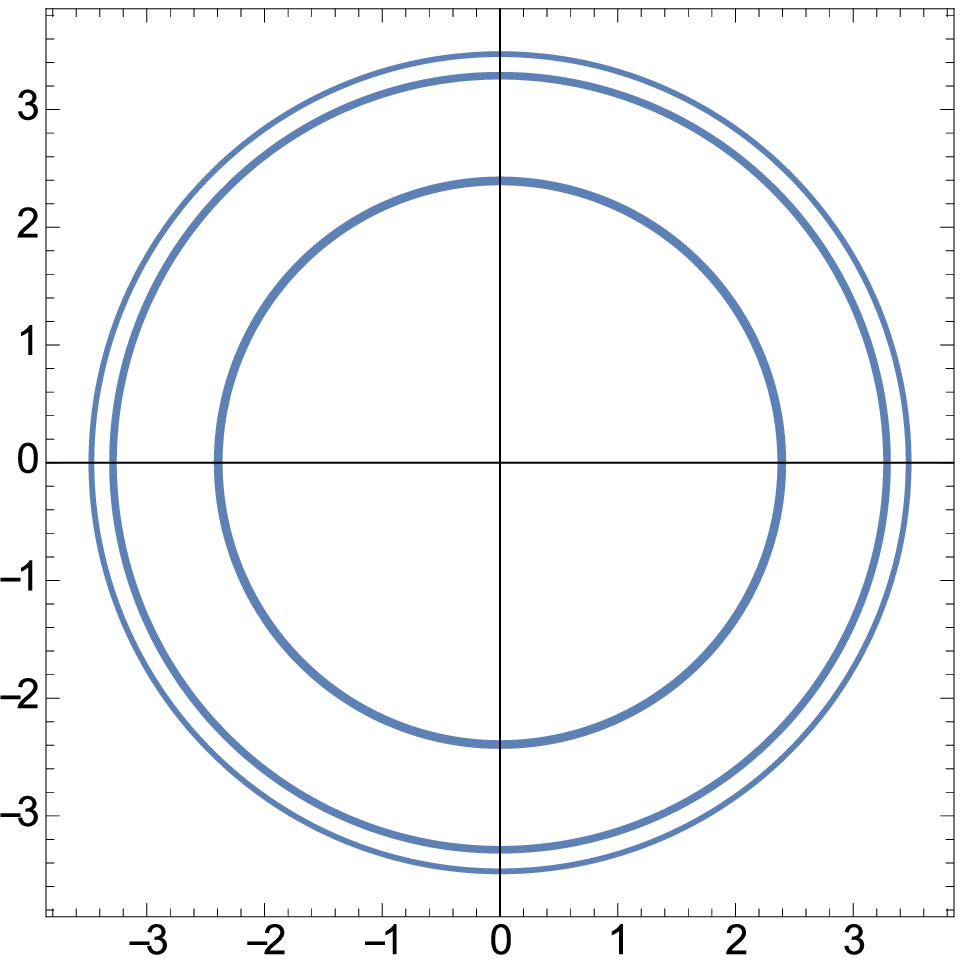}\label{fig:ring3}}
\subfigure[$x=0.9$]{\includegraphics[scale=0.55]{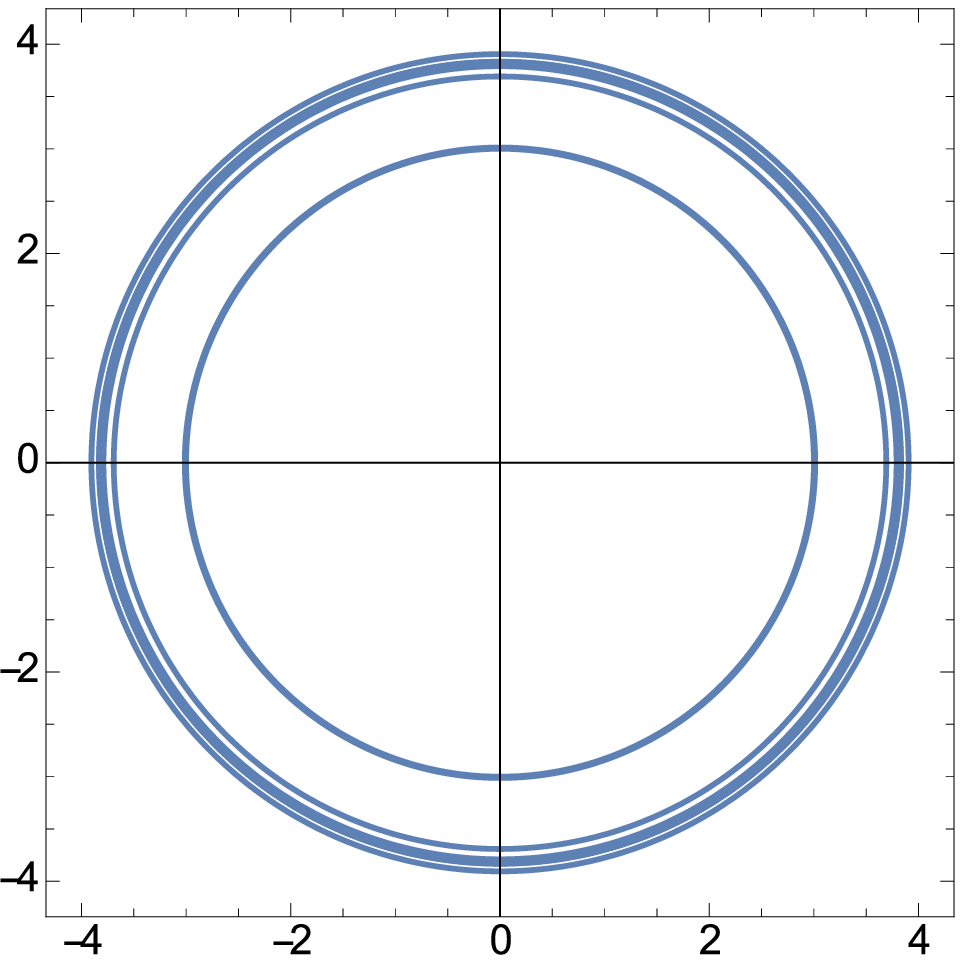}\label{fig:ring4}}
\subfigure[$x=0.93$]{\includegraphics[scale=0.55]{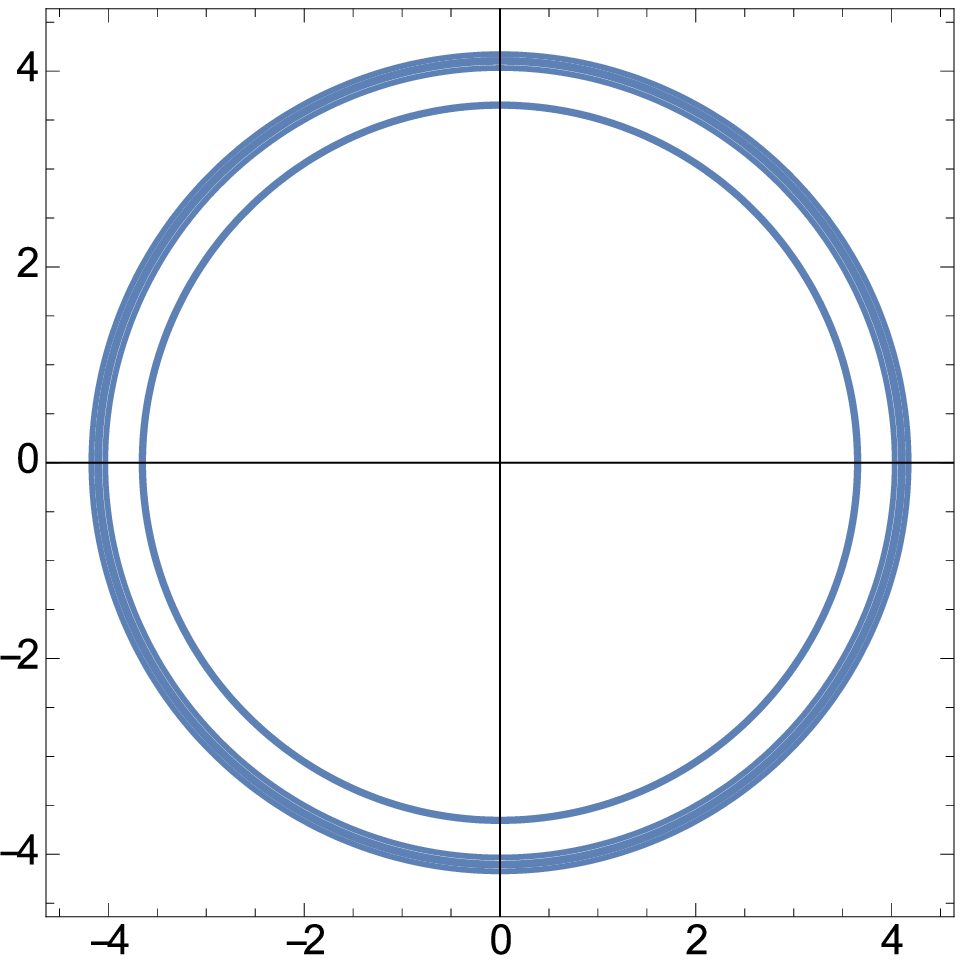}\label{fig:ring5}}
\subfigure[$x=0.96$]{\includegraphics[scale=0.55]{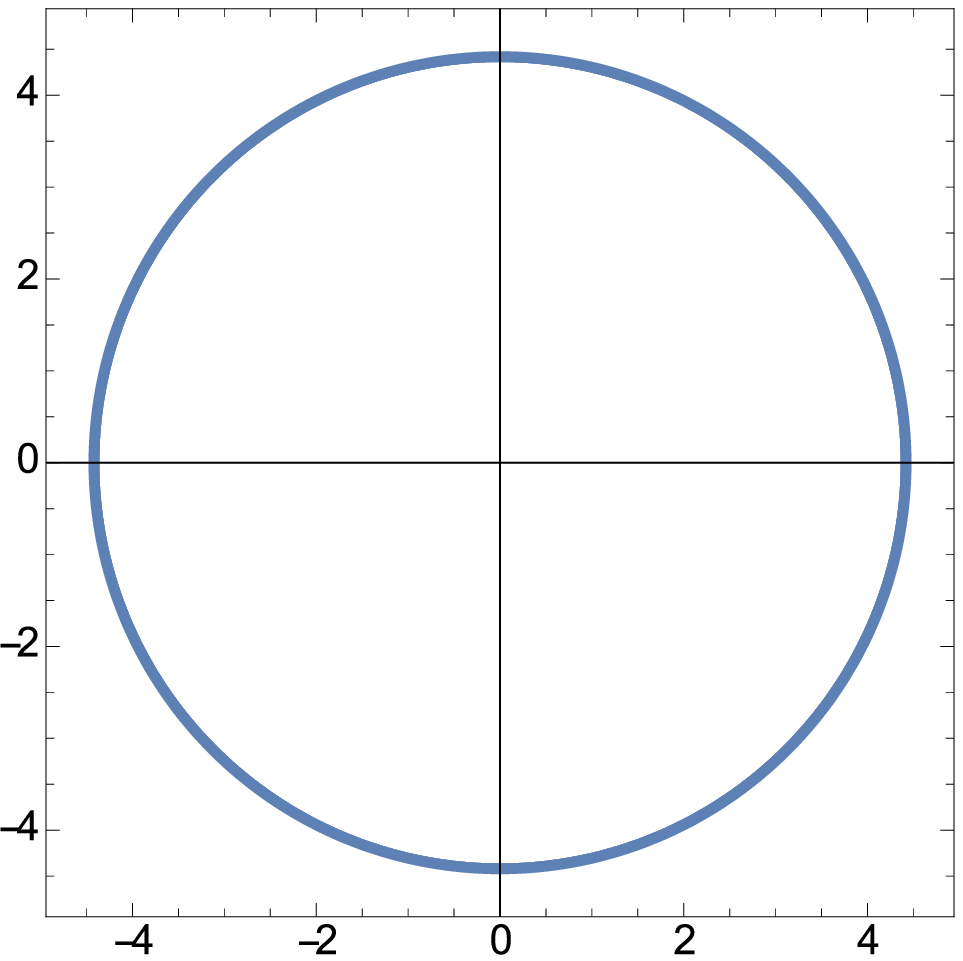}\label{fig:ring6}}
\subfigure[$x=1.1$]{\includegraphics[scale=0.55]{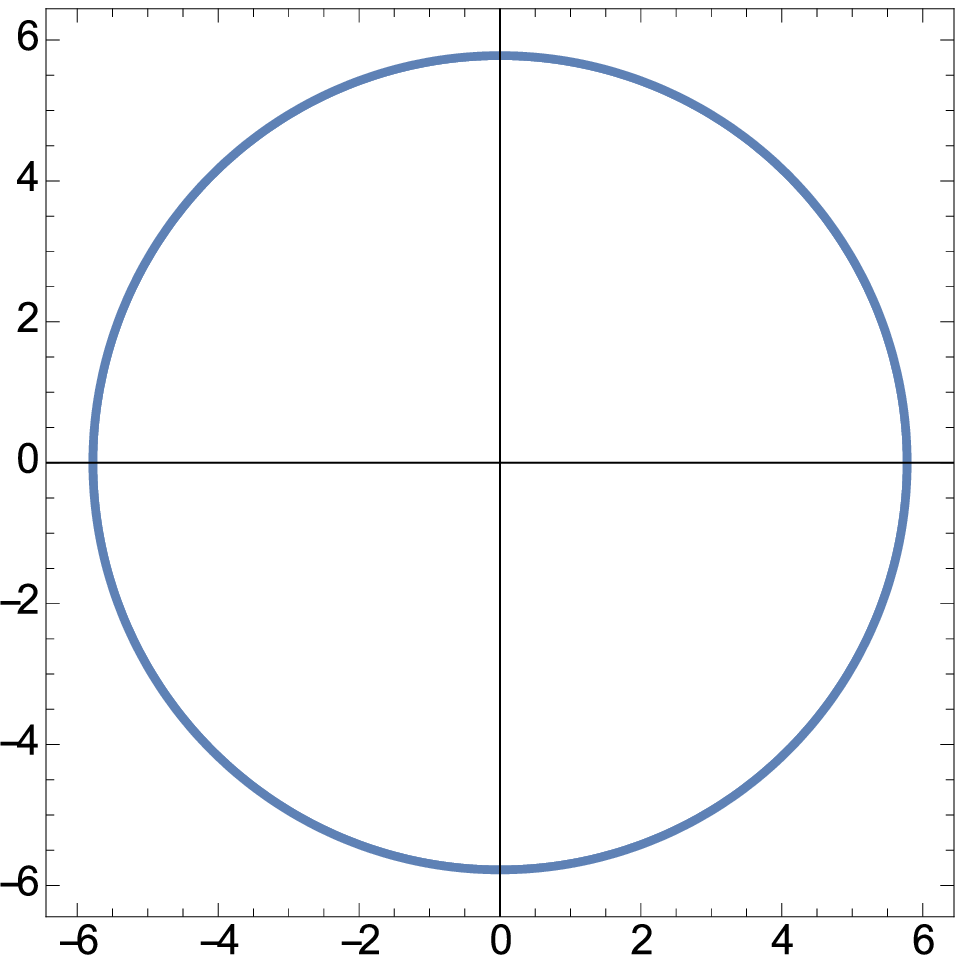}\label{fig:ring7}}
\caption{The relativistic Einstein rings for different $x$ ($=\frac{r_0^2}{|\kappa|}$) and $\kappa=-1$.}
\label{fig:ring}
\end{figure}

To study the relativistic images formed due to strong gravitational lensing of light coming from a distant light source, we will, for simplicity, assume that the observer, the lens (the wormhole or the black hole), and the distant point light source are all aligned. We also consider the
situation where the observer and the light source are far away from the lens. Therefore, in the observer’s sky, the relativistic images will be concentric rings (known as relativistic Einstein rings) of radii given by the corresponding impact parameters $b(r_{tp})$. These impact parameter values $b(r_{tp})$ can be obtained by solving $\alpha\simeq 2\pi n$, where $n$ is the ring number \citep{SL3}. Figure \ref{fig:ring} shows the relativistic Einstein rings in the observer's sky. Note that there are two sets of infinite number of relativistic Einstein rings in the case when there are two photon spheres (effective photon sphere at the throat and the outer photon sphere). This is something different from the black hole case (Fig. \ref{fig:ring7}) where there is only one set of relativistic rings formed due to a photon sphere. 

Also note that, when there is only one photon sphere (effective photon sphere or the outer photon sphere), the rings are closely clumped together. But, when there are two photon spheres, the rings at the outer photon sphere are less closely packed than those at the inner photon sphere. Here we would like to point out that photons with impact parameters less than the critical value corresponding to the inner photon sphere will get captured by the wormholes and travel to the other side. Therefore, if no radiation or little amount of radiation comes from the other side, then such a wormhole will cast a shadow \cite{WL17}, with the inner set of the relativistic images forming the outer bright edge of the shadow.

\section{Conclusion}
\label{sec:conclusion}
Horizonless compact objects with light rings (or photon spheres) are becoming increasingly popular, as these objects can be potential alternatives to black holes and can be laboratories for testing modifications to general relativity at the horizon scale. It is therefore important to study various properties of these objects and see whether or how we can distinguish these objects from black holes through observations. To this end, strong gravitational lensing, where a photon sphere plays a crucial role, is an important observational tool for probing the spacetime geometries around these objects. 

In this work, we have shown that spherically symmetric, static wormholes of the Morris-Thorne class can posses two photon spheres. In particular, we have shown that, in addition to a photon sphere present outside a wormhole throat, the throat can also act as an effective photon sphere. Such wormholes exhibit two sets of relativistic Einstein ring system formed due to strong gravitational lensing of light. If such type of wormhole casts a shadow at all, then the inner set of the relativistic Einstein rings will form the outer bright edge of the shadow. We have illustrated this further through a specific example. The above-mentioned rich and novel strong lensing feature points out the vital role that a wormhole throat can play. Such a novel lensing feature, which has not been pointed out before, may be useful in detecting wormholes in future observations.

\end{document}